\documentclass[twocolumn,letter]{jpsj3}

\usepackage{color}
\usepackage{graphicx}
\usepackage{dcolumn}
\usepackage{bm}

\begin{document}

\title{Relationship between Superconductivity and Antiferromagnetism in LaFe(As$_{1-x}$P$_{x}$)O Revealed by $^{31}$P-NMR}

\author{Shunsaku~Kitagawa$^{1,2,}$\thanks{E-mail address: shunsaku@crystal.kobe-u.ac.jp}$^,$\thanks{Present Address: Department of Physics, Kobe University, Kobe 657-8501, Japan}, Tetsuya~Iye$^{1,2}$, Yusuke~Nakai$^{1,2,}$\thanks{Present Address: Grad. Sch. of Sci, Tokyo Metropolitan Univ., Hachioji, Tokyo 192-0397, Japan}, Kenji~Ishida$^{1,2}$, Cao~Wang$^3$, Guang-Han~Cao$^3$, and Zhu-An~Xu$^3$}
\inst{$^1$Department of Physics, Graduate School of Science, Kyoto University, Kyoto 606-8502, Japan \\
$^2$Transformative Research Project on Iron Pnictides (TRIP), Japan Science and Technology Agency (JST), Chiyoda, Tokyo 102-0075, Japan\\
$^3$Department of Physics, Zhejiang University Hangzhou 310027, China}

\abst{We performed $^{31}$P-NMR measurements on LaFe(As$_{1-x}$P$_{x}$)O to investigate the relationship between antiferromagnetism and superconductivity.
The antiferromagnetic (AFM) ordering temperature $T_{\rm N}$ and the moment $\mu_{\rm ord}$ are continuously suppressed with increasing P content $x$ and disappear at $x = 0.3$ where bulk superconductivity appears.
At this superconducting $x = 0.3$, quantum critical AFM fluctuations are observed, indicative of the intimate relationship between superconductivity and low-energy AFM fluctuations associated with the quantum-critical point in LaFe(As$_{1-x}$P$_{x}$)O. The relationship is similar to those observed in other isovalent-substitution systems, e.g., BaFe$_{2}$(As$_{1-x}$P$_{x}$)$_{2}$ and SrFe$_{2}$(As$_{1-x}$P$_{x}$)$_{2}$, with the ``122'' structure.
Moreover, the AFM order reappears with further P substitution ($x > 0.4$). 
The variation of the ground state with respect to the P substitution is considered to be linked to the change in the band character of Fe-3$d$ orbitals around the Fermi level.
}

\abovecaptionskip=-5pt
\belowcaptionskip=-10pt
\setlength{\textheight}{660pt}

\maketitle

The interplay between superconductivity and magnetism is one of the hottest topics in condensed matter physics.
It has been believed that superconductivity is mediated by antiferromagnetic (AFM) fluctuations in cuprates and heavy-fermion superconductors\cite{T.Moriya_AP_2000,T.Moriya_RPP_2003,P.Monthoux_PRL_1991,C.Pfleiderer_RMP_2009}.
The relationship between superconductivity and AFM fluctuations in recent-discovered iron-based superconductors has also been discussed\cite{Y.Kamihara_JACS_2008,K.Ishida_JPSJ_2009,J.Paglione_Naturephys_2010,K.Kuroki_PRL_2008,J.Dai_PNAS_2009,Y.Nakai_PRL_2010,T.Iye_JPSJ_2012,T.Iye_PRB_2012,F.L.Ning_PRL_2010,S.Jiang_JPCM_2009,S.Kasahara_PRB_2010,Y.Nakai_PRB_2013}.
Up to now, it has been reported that superconductivity appears around the AFM quantum critical point in the Ba122 system, indicative of the strong relationship between superconductivity and low-energy AFM fluctuations\cite{Y.Nakai_PRL_2010,T.Iye_PRB_2012,F.L.Ning_PRL_2010,Y.Nakai_PRB_2013,S.Jiang_JPCM_2009,S.Kasahara_PRB_2010}.
On the other hand, superconductivity is linked not to the low-energy AFM fluctuations probed by NMR spectroscopy but to the local stripe spin correlation with $Q_{\rm stripe} = (\pi, 0)$ (unfolded Brillouin zone) in LaFeAs(O$_{1-x}$F$_{x}$)\cite{Y.Nakai_JPSJ_2008,Y.Nakai_NJP_2009,S.Kitagawa_PRB_2010,T.Nakano_PRB_2010}.
It seems that the roles of AFM fluctuations in superconductivity are different between the ``122'' and ``1111'' systems. 
However, in the ``1111'' system, most experiments have been performed on the electron-doped system. 
In addition, a different conclusion that low-energy AFM fluctuations probed by NMR spectroscopy are related to superconductivity in LaFeAs(O$_{1-x}$F$_{x}$) has been reported by other research groups\cite{T.Oka_PRL_2010}.
Therefore, in order to clarify the universal feature between superconductivity and AFM fluctuations in iron-based superconductors, experimental research on other ``1111'' systems is necessary.

In this study, we performed $^{31}$P-NMR measurements on P-substituted LaFe(As$_{1-x}$P$_{x}$)O ($x = 0.1, 0.2, 0.3, 0.4$, and 0.5) to investigate the relationship between superconductivity and antiferromagnetism.
Isovalent P-substitution does not introduce charge carriers and thus it corresponds to applying chemical pressure.
According to our NMR results, the AFM ordering temperature $T_{\rm N}$ and the moment $\mu_{\rm ord}$ are continuously suppressed with increasing P content $x$ up to 0.3 and 
bulk superconductivity appears when antiferromagnetism is suppressed.
In contrast to those observed in LaFeAs(O$_{1-x}$F$_{x}$)\cite{Y.Nakai_NJP_2009}, low-energy AFM fluctuations remain even at $x = 0.3$, where the superconducting transition temperature $T_{\rm c}$ is maximum, suggesting that the superconductivity is associated with low-energy AFM fluctuations with a quantum critical character, as observed in BaFe$_{2}$(As$_{1-x}$P$_{x}$)$_{2}$\cite{Y.Nakai_PRL_2010} and Ba(Fe$_{1-x}$Co$_{x}$)$_{2}$As$_{2}$\cite{F.L.Ning_PRL_2010}.
In addition, the AFM order reappears with further P-substitution.
We consider that a change in the band character around the Fermi level with P substitution induces a variation of the ground state in LaFe(As$_{1-x}$P$_{x}$)O.

\begin{figure*}[tb]
\vspace*{-10pt}
\begin{center}
\includegraphics[width=18cm,clip]{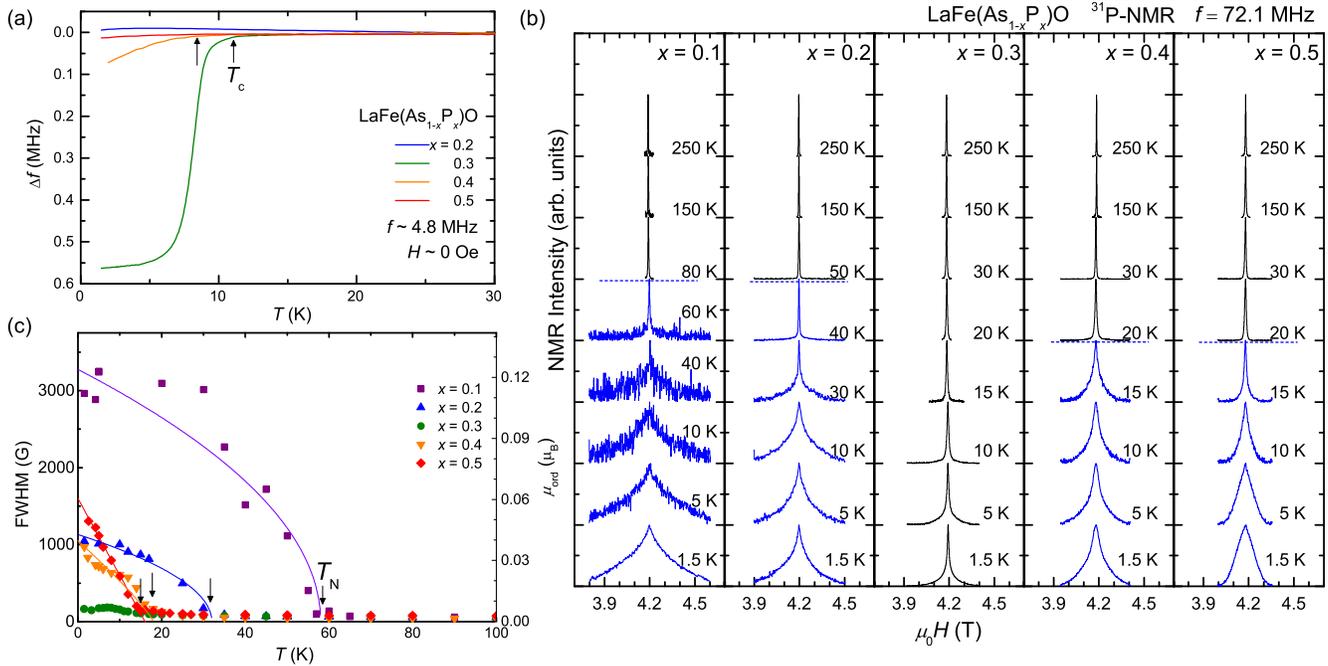}
\end{center}
\caption{(Color online) (a) Meissner signals for LaFe(As$_{1-x}$P$_{x}$)O ($x = 0.2 - 0.5$) measured using an NMR coil at $\mu_0 H = 0$~T. The arrow indicates $T_{\rm c}$. 
(b) Temperature dependence of $^{31}$P-NMR spectra at $f = 72.1$~MHz for $x = 0.1 - 0.5$.
(c) Temperature dependence of FWHM of $^{31}$P-NMR spectrum for $x = 0.1 - 0.5$. Below $T_{\rm N}$, FWHM rapidly increases with cooling, except for $x = 0.3$.
The derived $\mu_{\rm ord}$ values are shown on the right-hand axis.
The solid lines are visual guides.}
\label{Fig.1}
\end{figure*}

Polycrystalline samples of LaFe(As$_{1-x}$P$_{x}$)O ($x = 0.1- 0.5$) synthesized by solid-state reaction were ground into powder for NMR measurements\cite{C.Wang_EPL_2009}. 
$T_{\rm c}$ is determined from the Meissner signal measured using an NMR coil, as shown in Fig.~\ref{Fig.1}(a).
The observed temperature dependence of the Meissner signal is evidence of bulk superconductivity occurring at $T_{\rm c}$ = 11~K when $x$ = 0.3, weak (nonbulk) superconductivity at 8~K when $x$ = 0.4, and the lack of superconductivity for the other samples. 
These results are consistent with the previous report\cite{C.Wang_EPL_2009}.
A conventional spin-echo technique was utilized for the following NMR measurement.
The AFM ordering temperature $T_{\rm N}$ is determined from the peak of the nuclear spin-lattice relaxation rate divided by the temperature $1/T_1T$ and the increase in the NMR linewidth.  

First, we focus on the evolution of an ordered moment $\mu_{\rm ord}$ upon P-substitution through the $^{31}$P-NMR spectrum.
Figure \ref{Fig.1}(b) shows the $H$-swept $^{31}$P-NMR spectra at various temperatures and $x$ values ($x = 0.1 - 0.5$) measured at 72.1~MHz.
All $^{31}$P-NMR spectra consist of a single and almost isotropic line shape, as expected for an $I = 1/2$ nucleus.
The linewidth of the spectrum, except for $x = 0.3$, increases significantly below $T_{\rm N}$ while the peak position of each spectrum does not change very much, as shown in Fig.~\ref{Fig.1}(b).
In the AFM state of iron pnictides, it was reported that Fe ordered moments lying in the $ab$ plane with stripe correlations induce an internal magnetic field $H_{\rm int}$ along the $c$-axis at the As and P sites owing to the off-diagonal term of the hyperfine coupling tensor\cite{Cruz_Nature_2008,K.Kitagawa_JPSJ_2008}.
In such a commensurate stripe-type AFM ordered state, the powder pattern of $I = 1/2$ becomes nearly rectangular\cite{H.Kinouchi_PRB_2013}.
However, the obtained spectra show a Lorentzian-like shape, indicative of the distribution of $H_{\rm int}$.
Such an $H_{\rm int}$ distribution can be interpreted in terms of the incommensurability of AFM order or the distribution of the amplitude of $\mu_{\rm ord}$.

\begin{figure}[tb]
\vspace*{-60pt}
\begin{center}
\includegraphics[width=9cm,clip]{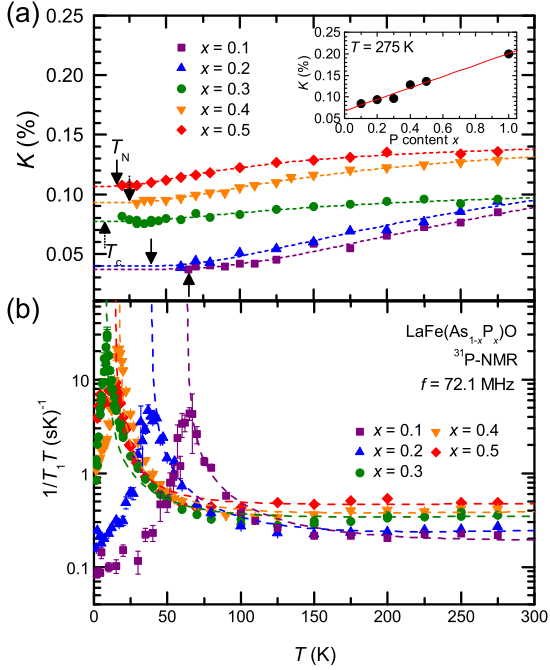}
\end{center}
\vspace*{-30pt}
\caption{(Color online) (a) (Main panel) Temperature dependence of Knight shift at $f = 72.1$~MHz for $x = 0.1 - 0.5$. The Knight shift slightly decreases upon cooling in all samples.
The broken lines are fitting lines. 
(Inset) P content $x$ dependence of the Knight shift at 275~K. The Knight shift at 275~K linearly increases with increasing $x$, suggesting a linear increase in the density of states.
The solid line is a visual guide.
(b) Temperature dependence of the spin-lattice relaxation rate $1/T_1T$ for $x = 0.1- 0.5$. $1/T_1T$ shows a peak at $T_{\rm N}$ corresponding to the critical slowing down at $x = 0.1, 0.2, 0.4$, and 0.5, which indicates magnetic ordering. On the other hand,  $1/T_1T$ at $x = 0.3$ drops sharply at $T_{\rm c}$ due to the opening of the superconducting gap.
The broken lines are fitted lines.}
\label{Fig.2}

\end{figure}

For simplicity, we use the FWHM to estimate the average $H_{\rm int}$, which is proportional to $\mu_{\rm ord}$.
Figure \ref{Fig.1}(c) shows the temperature dependence of FWHM of the $^{31}$P-NMR spectrum for all samples.
The FWHMs of the spectrum at 250~K are almost the same among the samples ($\sim$~50~G), indicating that the distribution of the bulk susceptibility is negligible.
The FWHM suddenly increases below $T_{\rm N}$, except for $x = 0.3$, representing the occurrence of internal magnetic fields.
At $x = 0.3$, FWHM slightly increases below 30 K, which is well above $T_c = 11$ K, but the increase is much smaller than those in the other samples. In addition, $1/T_1T$ does not show a maximum at 30 K, but shows a multicomponent behavior below 30 K, as will be discussed later.
Therefore, the $x = 0.3$ sample includes a small difference in $x$ concentration, and shows magnetic ordering due to the distribution of $x$.
The FWHM at 1.5~K continuously decreases with increasing $x$ up to 0.3.
With further increase in $x$, the AFM order reappears above $x = 0.4$.
In order to estimate $\mu_{\rm ord}$ from the internal field at the P site, $H_{\rm int}^{\rm P}$, we need to estimate the off-diagonal term of the hyperfine coupling tensor, $B_1$, at the P site, $B_1^{\rm P}$, since $\mu_{\rm ord}$ is approximately expressed as $H_{\rm int}^{\rm P} = B_1^{\rm P} \times \mu_{\rm ord}$. Using $B_1 = 4.4 {\rm T}/\mu_{\rm B}$\cite{S.Kitagawa_PRB_2010} at the As site and the ratio of $1/T_1$ at the As site to that at the P site measured in LaFe(As$_{0.7}$P$_{0.3}$)O (not shown), $B_1^{\rm P}$ is estimated to be $\sim 2.6~{\rm T}/\mu_{\rm B}$.
The derived temperature dependence of $\mu_{\rm ord}$ is shown on the right-hand axis in Fig. \ref{Fig.1}(c).

Figure \ref{Fig.2}(a) shows the temperature dependence of the Knight shift.
The Knight shift $K$ was determined from the peak field of the $^{31}$P-NMR spectrum. 
$K = 0$ was determined using the reference material H$_{3}$PO$_{4}$. 
$K$, which is a measure of the local susceptibility at the nuclear site, is described as $K = K_{\rm spin} + K_{\rm chem}$, where $K_{\rm spin}$ is the spin part of $K$ related to the uniform spin susceptibility $\chi(q = 0)$, which is proportional to the density of states at around the Fermi energy $N(E_{\rm F})$ in the paramagnetic state. In addition, $K_{\rm chem}$ is the chemical shift, which is generally temperature-independent and is assumed to be independent of $x$.
$K$ at 275~K is linearly proportional to $x$, as shown in the inset of Fig.~ \ref{Fig.2}(a), suggesting that $N(E_{\rm F})$ increases with increasing $x$.
$K$ in (La,Ca)FePO is also plotted as a reference for $x = 1$\cite{Y.Nakai_PRL_2008}.
Since P is isovalent with As, this increase in $N(E_{\rm F})$ originates from a change in the band structure induced by chemical pressure.
P substitution changes the band structure and $N(E_{\rm F})$ increases.
This is in contrast to that observed in BaFe$_{2}$(As$_{1-x}$P$_{x}$)$_{2}$\cite{Y.Nakai_PRL_2010,T.Iye_JPCS_2012}, where $K$ is almost independent of $x$ up to 0.64.
In all samples of LaFe(As$_{1-x}$P$_{x}$)O, $K(T)$ slightly decreases upon cooling, as observed in electron-doped systems\cite{K.Ahilan_PRB_2008,A.Kawabata_JPSJ_2008_2}.
These temperature dependences can be explained by the energy dependence of the density of states, as proposed by Ikeda\cite{H.Ikeda_JPSJ_2008}.
$K(T)$ is fitted using an activation type equation: $K(T) = a + b\exp(-\Delta/k_{\rm B}T)$, where $\Delta$ is the activation energy.
The $\Delta$ values are 373, 284, 156, 167, and 103~K for $x = 0.1, 0.2, 0.3, 0.4,$ and 0.5, respectively. 

Although the Knight shift gradually decreases upon cooling, $1/T_1T$ strongly increases toward $T_{\rm N}$ or $T_{\rm c}$, as shown in Fig.~\ref{Fig.2}(b).
$1/T_1$ was measured by a saturation recovery method.
Although the time dependence of the spin-echo intensity $M(t)$ after the saturation of nuclear magnetization can be fitted to a theoretical curve of the nuclear spin $I = 1/2$ with a single component of $T_1$ at high temperatures, $M(t)$ deviates from the theoretical curves with a single $T_1$ component and shows a multi-$T_1$-component behavior at low temperatures.  
Then, $T_1$ for all $x$ values in this paper was determined by fitting it to the stretched exponential function $[M(\infty) - M(t)]/M(\infty) =  c \exp[-(t/T_1)^{\beta}]$, where $c$ is the initial saturation of the nuclear magnetization and $\beta$ describes the homogeneity of $T_1$.
At high temperatures, $\beta \simeq 1$ in all samples and $\beta$ starts to decrease below 125, 100, 30, 40, and 50~K for $x = 0.1, 0.2, 0.3, 0.4$, and 0.5, respectively.
One of the reasons for the decrease in $\beta$ is the anisotropy of $1/T_1$, since $1/T_1$ becomes very anisotropic below the structural phase transition temperature in the iron pnictides\cite{K.Kitagawa_JPSJ_2011,Y.Nakai_PRB_2012}.
In our measurements, $1/T_1$ includes magnetic fluctuations along all directions since powder samples were measured.
Another possibility is the distribution of $x$. In the $x$ region where $T_{\rm N}$ significantly changes with $x$, the tiny distribution of $x$ would induce a multicomponent behavior in the recovery of $M(t)$, although $T_{\rm N}$ is clearly determined. 
Therefore, $T_1$ determined with the stretched exponential function is regarded as the average $T_1$ with respect to $x$.   
$1/T_1T$ is expressed by the wave-vector $\bm{q}$-integral of the imaginary part of the dynamical susceptibility $\sum_{\bm{q}} \chi"(\bm{q}, \omega)$.
Therefore, the Curie-Weiss-like enhancement of $1/T_1T$ shown in Fig.~\ref{Fig.2}(b) and the gradual decrease in the Knight shift related to $\chi(\bm{q} = 0)$ in the normal state indicate the development of low-energy AFM ($\bm{q} \neq 0$) fluctuations.
Upon further cooling, $1/T_1T$ shows a peak at $T_{\rm N}$ corresponding to the critical slowing down at $x = 0.1, 0.2, 0.4$, and 0.5, which indicates magnetic ordering.
$T_{\rm N}$ is unambiguously determined by the peak of $1/T_1T$.
On the other hand,  $1/T_1T$ at $x = 0.3$ drops sharply at $T_{\rm c}$ due to the opening of a superconducting gap.

\begin{figure}[tb]
\vspace*{-10pt}
\begin{center}
\includegraphics[width=9cm,clip]{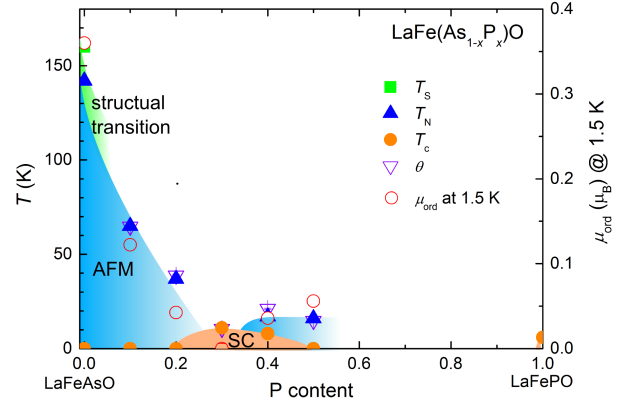}
\end{center}
\caption{(Color online) $T-x$ phase diagram for LaFe(As$_{1-x}$P$_{x}$)O.
Filled circles, square, and triangles indicate $T_{\rm c}$, structural phase transition temperature $T_{\rm S}$, and $T_{\rm N}$, respectively.
The open circles and triangles indicate $\mu_{\rm ord}$ at 1.5~K and $\theta$ estimated from the fitting of the temperature dependence of $1/T_1T$, respectively.
$\mu_{\rm ord}$ = 0.36 $\mu_{\rm B}$ at $x = 0$\cite{Cruz_Nature_2008} is also plotted.}
\label{Fig.3}
\end{figure}

We fit the observed $1/T_1T$ to the equation in the phenomenological two-component model, $1/T_1T = (1/T_1T)_{\rm inter} + (1/T_1T)_{\rm intra}$. 
Here, $(1/T_1T)_{\rm inter} = C/(T + \theta)$, where $C$ is a constant and $\theta$ is the Weiss temperature, corresponds to the contribution of the interband two-dimensional AFM fluctuations expected in the self-consistent renormalization (SCR) theory, and $(1/T_1T)_{\rm intra} = d + e\exp(-\Delta/k_{\rm B}T)$ corresponds to the intraband contribution, which is proportional to $N(E_{\rm F})^2$.
We assume that $N(E_{\rm F})$ shows an activation-type temperature dependence as well as the Knight shift.
The activation energies estimated from the Knight shift are used for the fitting.
In addition, we assume that the values of the constant $C$ are the same among all samples, as observed in BaFe$_{2}$(As$_{1-x}$P$_{x}$)$_{2}$\cite{Y.Nakai_PRL_2010}.
Then, we can fit $1/T_1T$ using the above equation, as shown in Fig.~\ref{Fig.2}(b).
The obtained $\theta$ plotted in Fig. \ref{Fig.3} is almost the same as $T_{\rm N}$ and approaches 0~K at approximately $x~=~0.3$. 
This behavior suggests that superconductivity is strongly related to AFM fluctuations associated with the quantum-critical point, similarly to that observed in other isovalent-substitution systems BaFe$_{2}$(As$_{1-x}$P$_{x}$)$_{2}$ and SrFe$_{2}$(As$_{1-x}$P$_{x}$)$_{2}$.
This strong relationship between superconductivity and AFM fluctuations is in contrast to that observed in LaFeAs(O$_{1-x}$F$_{x}$)\cite{Y.Nakai_NJP_2009,T.Nakano_PRB_2010}.
The $1/T_1T$ of LaFeAs(O$_{1-x}$F$_{x}$) is suppressed markedly with F doping, whereas $T_{\rm c}$ does not change very much, suggesting a weak correlation between superconductivity and low-energy spin fluctuations probed by NMR spectroscopy.
In addition, the phase diagram of LaFe(As$_{1-x}$P$_{x}$)O is quite different from that of LaFeAs(O$_{1-x}$F$_{x}$) as shown below.
In LaFe(As$_{1-x}$P$_{x}$)O, the AFM order is continuously suppressed with P substitution, whereas the AFM order suddenly disappears with the first-order-like transition against F content in LaFeAs(O$_{1-x}$F$_{x}$)\cite{Q.Huang_PRB_2008,H.Luetkens_NatMat_2009,Y.Nakai_PRB_2012}.
These experimental results suggest that there are at least two pairing mechanisms in iron-based superconductors.

Finally, we summarize our NMR data of $T_{\rm N}$, $\mu_{\rm ord}$ at 1.5~K, and $\theta$ in the $T-x$ phase diagram of LaFe(As$_{1-x}$P$_{x}$)O, as shown in Fig.~\ref{Fig.3}.
The AFM order is suppressed with P substitution up to $x = 0.3$, but reappears above $x = 0.4$.
In the low-P AFM state ($0 \leq x < 0.3$), $T_{\rm N}$ and $\mu_{\rm ord}$ decreases from 135~K at $x = 0$ to $\sim$~0~K at $x = 0.3$, whereas $T_{\rm N}$ seems to be constant against P substitution in the high-P AFM state ($x \geq 0.4$).
Moreover, the trends of the AFM order seem to be different between the two AFM states, since $\mu_{\rm ord}$ in the low-P AFM state grows, following the mean-field-type dependence [$H_{\rm int}(T) \propto (T_{\rm N} - T)^{0.5}$], but $\mu_{\rm ord}$ increases linearly against $T$ in the high-P AFM state.
The high-P AFM state might be a short-range order such as a spin-glass state.
Therefore, to understand the nature of the new AFM state, neutron scattering measurements and thermodynamic measurements, such as specific heat measurements, are desired.
These differences might be explained by the features of  the nesting condition.
According to the band calculations, the $d_{x^2-y^2}$ orbital mainly contributes to the hole Fermi surfaces (FSs) at the $(\pi, \pi)$ point in the unfolded Brillouin zone, and the nesting between the hole and electron FSs is enhanced in the antiferromagnet LaFeAsO, while the $d_{3z^2-r^2}$ orbital mainly contributes in the paramagnet LaFePO\cite{V.Vildosola_PRB_2008,K.Kuroki_PRB_2009}.
This indicates that the P substitution changes the orbital characteristics of the hole FSs as well as the nesting properties, which induces AFM fluctuations.
Our NMR results suggest that AFM fluctuations become enhanced again at approximately $x = 0.5$, where the band character at the $(\pi, \pi)$ point is replaced. 
This might be a characteristic feature in LaFe(As$_{1-x}$P$_{x}$)O since such an AFM state has never been reported.
Therefore, LaFe(As$_{1-x}$P$_{x}$)O is a good system for studying the relationship between superconductivity and antiferromagnetism induced by nesting.

In conclusion, we found that $T_{\rm N}$ and $\mu_{\rm ord}$ are continuously suppressed with P substitution up to $x = 0.3$, where bulk superconductivity appears, similar to that observed in other isovalent-substitution systems.
With further P substitution, the AFM order reappears and the nature of the high-P AFM state seems to be different from that of the low-P AFM state.
We consider that the variation of the ground state with respect to P substitution is related to the change in the band character around ($\pi, \pi$) in the unfolded Brillouin zone.

\section*{Acknowledgments}
We thank H. Ikeda, S. Yonezawa, and Y. Maeno for support in the experiments and valuable discussions. 
This work was partially supported by Kyoto University LTM Center, by a ``Heavy Electrons'' Grant-in-Aid for Scientific Research on Innovative Areas  (No. 20102006) from The Ministry of Education, Culture, Sports, Science, and Technology (MEXT) of Japan, by a Grant-in-Aid for the Global COE Program ``The Next Generation of Physics, Spun from Universality and Emergence'' from MEXT of Japan, and by a Grants-in-Aid for Scientific Research from Japan Society for the Promotion of Science (JSPS), KAKENHI (S and A) (Nos. 20224008 and 23244075). 
SK is financially supported by a JSPS Research Fellowship.

\end{document}